\documentclass[aps,prc,twocolumn,tightenlines,nofootinbib,letter,showpacs]{revtex4-2}
\usepackage{dcolumn}
\usepackage{xcolor}
\usepackage{graphicx}
\usepackage{rotating}
\usepackage{natbib}
\usepackage{amsmath}
\usepackage{braket}
\renewcommand{\vec}[1]{\mbox{\boldmath$#1$\unboldmath}}
\begin{document}

\title{Cross sections of removal reactions populating weakly-bound residual nuclei}
\author{J.\,A. Tostevin}
\affiliation{Department of Physics, Faculty of Engineering and Physical Sciences,
University of Surrey, Guildford, Surrey GU2 7XH,
United Kingdom}
\date{\today}

\begin{abstract}
In many instances, single nucleon removal reactions from neutron-proton asymmetric
projectile nuclei populate final states in the residual nuclei that are very weakly
bound. Familiar examples include neutron removal reactions from neutron-rich
$^{11}$Be and $^{12}$Be, the latter populating the well-known $1/2^+$ halo
ground-state and $1/2^-$ excited-state of $^{11}$Be - both states less than 1 MeV
from the first neutron-decay threshold. Numerous additional examples arise in reactions
of asymmetric $p$- and $sd$-shell nuclei. The importance of this weak residue binding
upon calculated single-nucleon removal reaction cross sections is quantified by means
of model calculations that neglect or include the dissociation degree of freedom of
the residual nuclei. The calculated removal-reaction cross sections for two
representative $p$-shell projectiles indicate that an explicit treatment of
these residue break-up effects is unnecessary and that the differences between
the break-up and no break-up calculations are small provided a consistent description
of the residue structure and density is used.
\end{abstract}
\maketitle

\section{Introduction}
The study of direct reactions in which a single nucleon is removed from a projectile nucleus continue
to be instrumental in extracting spectroscopic information for exotic nuclei. Specifically, they
contribute to the study of single-particle degrees of freedom and help identify the active
valence single-particle orbitals near the Fermi-surfaces, their ordering and their occupancies. A primary motivation of
such direct-reaction experiments and analyses is to infer quantitative spectroscopic
information on rare, neutron-proton asymmetric nuclei. Extracting such information
relies upon the ability of the combined direct-reaction and nuclear structure models
used to calculate absolute cross sections with a fair degree of reliably. Whether the
direct reactions used are intermediate-energy nucleon removal (knockout) reactions,
the mechanism of interest here, or lower-energy single-nucleon transfer reactions,
each projectile initial-state to residual nucleus final-state transition is
quantified by comparing the measured final-state partial cross section with that
computed with a direct-reaction model that: (a) includes sufficiently accurately
(but usually approximately) the dominant reaction mechanisms, and (b) includes an
initial- to final-state single-particle overlap function obtained from appropriate
nuclear structure model calculations.

In the case of intermediate-energy single nucleon removal reactions, the eikonal
direct-reaction dynamical model considered here, for details see for example Ref. \cite{Han03} and
references therein, has, typically, been combined with nuclear shell-model overlaps:
that is the spectroscopic factors together with assumed radial form-factors of the
removed nucleons \cite{TOS14}. Within this methodology, for reactions of a mass
$A$ projectile with a target of light nuclei, usually Be or C, all details of the
interaction of the residual, mass $A$$-$1 nucleus with the target are described
by its eikonal-model elastic S-matrix, $S_{{r}}^{}(b_r)$, computed as a function
of their collision impact parameter $b_r$. It is this input to the reaction dynamical
model that carries information on the size, structure and binding of the residual
nucleus through the refractive and absorptive nature of the residue-target
interaction.

In many cases, particularly when the projectile nuclei approach the drip-lines,
the residual nuclei and/or their excited states populated by nucleon removal
are only very weakly bound and so have spatially-extended wave functions. Examples
include the neutron removal reactions from neutron-rich $^{11}$Be and $^{12}$Be,
populating the $1^-$ and $2^-$ excited states of $^{10}$Be near 6 MeV \cite{Tom},
and the well-known $1/2^+$ halo ground-state and $1/2^-$ excited-state of $^{11}$Be
\cite{Navin}, respectively. Typically however, removal reaction
analyses \cite{TOS14} have exploited conventional optical limit (OL) eikonal-model
calculations for the residue-target S-matrix in which the residue structure is
represented by parameterised neutron and proton one-body density functions - usually
derived from Hartree-Fock calculations, empirical information on nuclear mass
and/or charge radii, or from $(A,Z)$ size systematics. This density-based approach
neglects the break-up degree of freedom of the populated residue final states.

An early mention of the possible effects of these break-up channels upon the
residue-target elastic S-matrix \cite{Lewes} suggested an increased residue-target
absorption - leading to a reduced residue survival probability and a reduced
removal-reaction cross section. However, in that discussion, these derived cross
section reductions
were computed relative to no break-up, OL calculations that used only a simple (Gaussian)
parameterised residue density and not the {\em actual} density computed consistently
from the structure of the weakly-bound residue - information that was included in the
comparison few-body calculations that included break-up degrees of freedom.

We present a more considered assessment of the importance of such break-up of
weakly-bound residual nuclei on calculated single-nucleon removal reaction cross
sections. Here, simplified but realistic model calculations that: (i) neglect, and
(ii) include the break-up degree of freedom of the residue, and that treat the
residue structure in a consistent way, are compared. The $^9$C($-p$)
and $^{10}$C($-n$) reactions, which involve removal of a weakly-bound proton, $S_p=
1.3$ MeV, and a strongly-bound neutron, $S_n=21.3$ MeV, respectively, and that
populate the weakly-bound $^8$B and $^9$C residues, with proton separation energies
0.137 MeV and 1.3 MeV, are used as representative test cases. Since only the ground
states of these residues are particle bound, inelastic excitations other than
break-up do not raise complications. We show that when the two sets of calculations are
performed consistently, the computed nucleon removal cross sections when neglecting
and including break-up degrees of freedom of the residues are essentially identical.
An earlier, shorter version of this discussion was the subject of the supplementary
material to a recent study of several $p$-shell mirror nuclei \cite{p-shell}.

\begin{figure}[b]
\vspace{-0.3cm}
\begin{center}
\includegraphics[width=0.90\columnwidth,angle=0,scale=1]{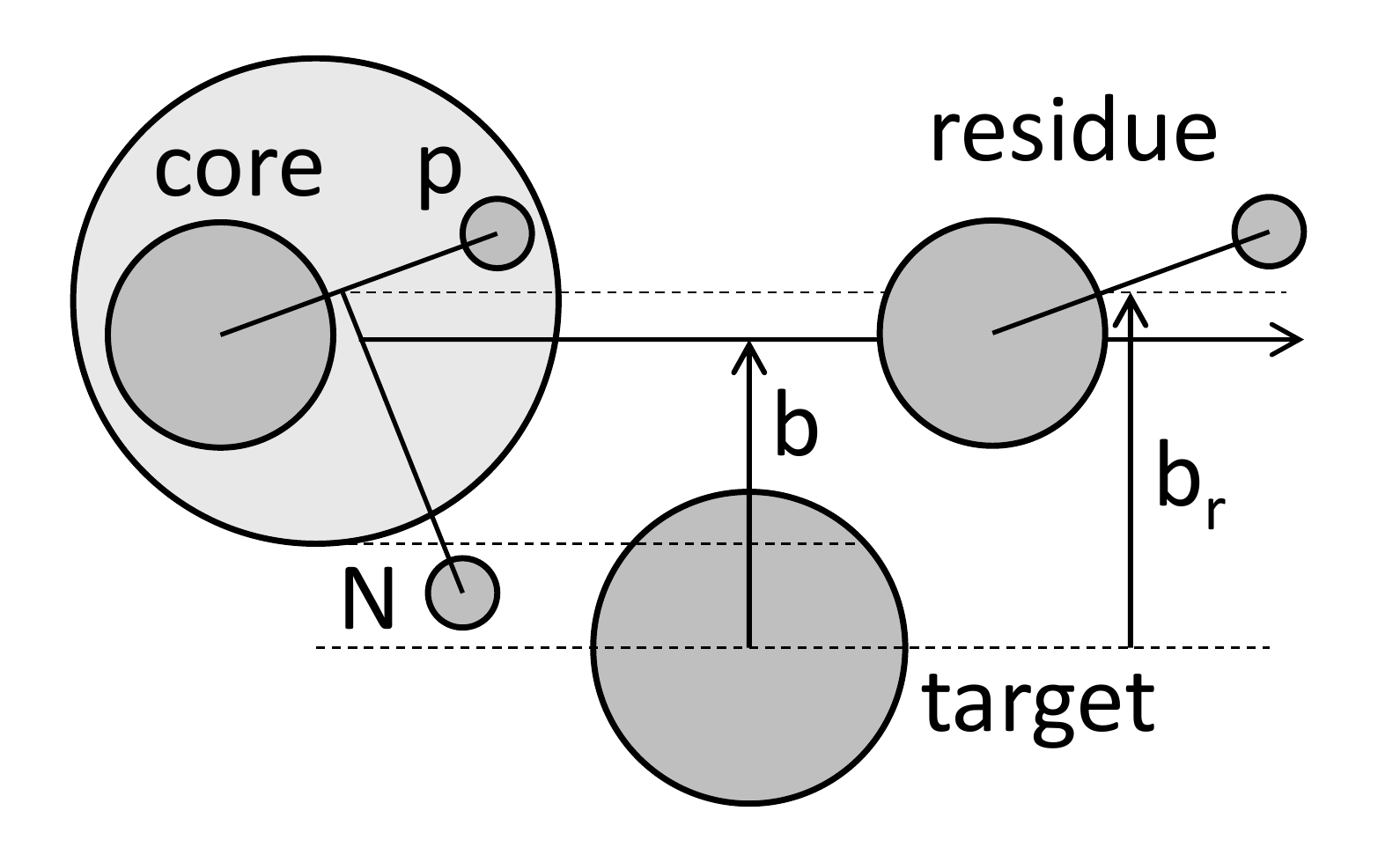}
\end{center}\vspace{-0.4cm}
\caption{Schematic of a sudden, nucleon (N) removal reaction at a projectile
centre-of-mass (cm) impact parameter $b$, leaving a weakly-bound, two-body (core$+p$)
residual nucleus with cm impact parameter $b_r$. In this few-body description, and for
the spatial configuration shown, the core and $p$ constituents of the bound residue
do not overlap and interact strongly with the target and will be transmitted. On the other
hand, within a density-based description, with no reference to the orientation of the
core$+p$ system, the one-body density of the residue (light-shaded circle) will overlap
the target at this $b$, resulting in a degree of absorption and reduced transmission.}
\label{fig:one}
\end{figure}

\section{Few-body considerations}
We consider the break-up degrees of freedom of weakly-bound residual nuclei within
a few-body, eikonal-model approximation. In the calculations presented, the $^8$B
and $^9$C reaction residues are treated as weakly-bound two-body (core + proton)
systems and we neglect any explicit consideration of the internal degrees of freedom
of these core nuclei, describing these using Gaussian densities with appropriate root
mean squared (rms) radii. Effects of the structure of these residues on the eikonal-model
$^9$C($-p$) and $^{10}$C($-n$) removal cross sections involve the residue-target
elastic S-matrix. Here, these S-matrices are computed: (i) from the residue ground-state
density, the optical limit, which neglects core$+p$ break-up channels, and (ii) taking
account explicitly of the residue's two-body cluster structure, its extended wave
function, and its break-up. These two approaches were designated the static-density
and few-body approaches, respectively, in the composite projectile reaction cross
sections work of Refs. \cite{AT} and \cite{ATT}.

The inclusion of break-up, approach (ii), does not necessarily imply the residue-target
S-matrix is more absorptive and the survival of the weakly-bound residue less likely
-- as the extended nature of the wave function of the residue affects the spatial
overlap of its constituents with the target, see e.g. Refs. \cite{AT,ATT} and Fig. 1.
At beam energies of order 800 MeV/nucleon \cite{AT,ATT}, where the interactions with
the target are overwhelmingly absorptive, including the effects of break-up of extended
two- and three-body systems was actually shown to reduce their calculated reaction cross
sections, $\sigma_R$, compared to calculations using the no break-up, density-based
OL model \cite{ATT}. Overall, the explicit treatment of the few-body nature of the
collision found it to be more transparent and less absorptive. This result was shown
by Johnson and Goebel \cite{JG} to be a very general result for strongly-absorbed
systems, and was also discussed using a simplified, semi-analytic model in
Ref. \cite{NN}. The reason for this reduced absorption, see Fig. 1, was that in many
spatial configurations of the separated constituents they will not overlap with or
interact strongly with the target. The additional transparency of the collision with
the light target nucleus, due to the granular, cluster nature of the extended few-body
system, was found to more than compensate for any additional absorption due to the
removal of flux from the elastic channel into break-up channels.

Of course, the eikonal-model stripping and diffraction dissociation components of
the nucleon removal cross-sections that produce the residues have a more complicated
dependence on the residue-target S-matrix than simply its total absorptive content --
as determined by its reaction cross section $\sigma_R$. These components are given
explicitly in Eqs.\ (2) and (4) of Ref. \cite{NS2k}. In such removal reaction calculations,
both the radial extent and geometry of the residue wave-function and of its density
(principally its rms radius \cite{AG2008}) determine the absorption profile of the
S-matrix as a function of impact parameter. This, in turn, determines which parts
of the bound-state wave function of the removed nucleon are sampled and thus
contribute to the cross section. Here, we quantify these residue structural effects
upon the calculated nucleon removal cross-section at intermediate energies of order
100 MeV/nucleon.

\section{Density-based S-matrices}
For the no break-up, density-based OL calculations, case (i) above, we require the neutron
and proton single-particle densities of the residue. Here we compute these from the
two-body, bound-state relative motion wave function, ${\Phi}_{0}^{}$, of its core ($c$)
and valence nucleon ($v$) and the internal densities of these constituents. The
two-body model
residue density can be written
\begin{eqnarray}
\rho_{r}(r) = \hat{\rho}_{{c}}(r) + \hat\rho_v(r)~~,\label{dddens}
\end{eqnarray}
where $\hat{\rho}_{c}(r)$ and $\hat\rho_v(r)$ are the contributions from the core
and valence nucleon in the residue centre-of-mass (cm) frame \cite{ATT}. Assuming
that the free core has internal density ${\rho}_{c}(r)$, then $\hat{\rho}_{c}(r)$ is
obtained by folding this intrinsic density with $\rho_{\text{cm}}({r})$, the
distribution of the motion of the cm of the core within the residue
bound state. Thus,
\begin{eqnarray}
\hat{\rho}_{c}(\vec{r}\,) = \int d\vec{x} \ \rho_{c}(\vec{r} - \vec{x})
\,\rho_{\text{cm}}(\vec{x})\ ,
\end{eqnarray}
where the bound-state-generated core cm distribution is
\begin{eqnarray}
\rho_{\text{cm}}(\vec{r}\,)=A_r^3 \left\vert {\Phi}_{0} \left(A_r \vec{r}\,
\right)\right\vert^2~,
\end{eqnarray}
and $A_r\,$$=\,$$A-1$ is the residue mass number. The valence nucleon density relative
to the cm of the residue is
\begin{eqnarray}
\hat\rho_v(\vec{r}\,)=\left[\frac{A_r}{A_r-1}\right]^3 \left\vert {\Phi}_{0}
\left(\frac{A_r}{A_r-1} \vec{r}\right)\right\vert^2 ~.
\end{eqnarray}
In the calculations carried out here, the cores, of mass number $A_c=A_r-1=N_c+Z_c$,
are assumed to have a Gaussian density and the valence particle is a proton,
in which case the proton and neutron densities of the residue are
\begin{eqnarray*}
\rho_p(r) = [Z_c/A_c]\,\hat{\rho}_{c}(r) + \hat\rho_v(r),~~\rho_n(r)=
[N_c/A_c]\,\hat{\rho}_{c}(r)~.
\end{eqnarray*}

When calculating the eikonal S-matrix for each particle $x$ with the target from
their densities: as is used here (a) in case (i) for the interaction of the
residues, based on their proton and neutron densities, given above, and (b)
in case (ii) for the interactions of the residue's core and the proton with the
target, these are computed in the optical limit (or $t \rho$ and $t \rho\rho$
folding) approximation to their optical potential, ${\cal U}_{xt}$. Given these
complex optical potentials, each fragment $x$-target S-matrix, $S_{x}$, is
\begin{eqnarray}
S_{x}(b)=\exp \left[i {\cal O}_{xt}(b) \right]\,,
\end{eqnarray}
where the eikonal phase shift, ${\cal O}_{xt}(b)$$\,\equiv\,$$2\delta(b)$, as a
function of impact parameter, is given by the $R_3$ integral (the $z$-component
of $\vec{R}$ in the incident beam direction) through the optical interaction
at each impact parameter $b\equiv b_{xt}$. Specifically,
\begin{eqnarray}
{\cal O}_{xt}(b)=-\frac{1}{\hbar v}\,\int_{-\infty}^{\infty}\!\! dR_3\
{\cal U}_{xt} (\sqrt{b^2+R_3^2}) \ ,
\end{eqnarray}
where $v$ is the $x$-target relative velocity.

\section{Few-body model S-matrices}
For the few-body calculations, case (ii) above, that include break-up of the
residue, the eikonal elastic S-matrix of the two-body, composite residue is
constructed from the core and proton OL S-matrices, discussed above, as
\begin{eqnarray}
S_{{r}}^{}(b_r)=\langle \Phi_{0}| S_{c}(b_{c}) S_{p}(b_p)|\Phi_{0}
\rangle_{\text{spin}}\ .
\end{eqnarray}
That is, the S-matrix product, $S_{c}(b_{c}) S_{p}(b_p)$, that describes the
combined core and proton system at a fixed vector separation, $\vec{r}$, and
residue cm impact parameter, $b_r$, must be integrated
over all possible spatial configurations $\vec{r}$ weighted by the probability
of each configuration in the core-proton relative motion wave function $\Phi_{0}$.
This involves the residue ground-state position probability summed over the
nucleon and core spin variables, denoted $\langle\ \vert \Phi_{0}(\vec{r})\vert^2
\rangle_{\text{spin}}$, as in Ref.\ \cite{ATT}.

\section{Calculations and results}
We carry out model nucleon-removal calculations for the $^9$C($-p$) and $^{10}$C($-n$)
reactions at 100 MeV/nucleon on a $^9$Be target, representative of reaction  data
of topical interest \cite{p-shell}. The parameters used are as follows.

The $^7$Be and $^8$B cores of the $^8$B and $^9$C residual nuclei and the $^9$Be target
are described by Gaussian densities with rms radii of 2.31 fm, 2.38 fm and 2.36 fm,
respectively \cite{Ozawa}. Thus, with $g(\gamma,r)$ a normalised 3-dimensional
Gaussian function
\begin{eqnarray}
g(\gamma,r)=(\sqrt{\pi}\gamma)^{-3}\exp(-r^2/\gamma^2)~~,
\end{eqnarray}
the neutron and proton densities of these systems are taken to be
$\rho_p (r)= Z_x g(\gamma_x,r)$ and $\rho_n (r)= N_x g(\gamma_x,r)$,
where $x=c,t$, and with $\gamma$ values determined by the rms radii, given
\begin{eqnarray}
\langle r^2\rangle=3 \gamma^2/2\ .
\end{eqnarray}

From the proton and neutron one-body densities of the target ($t$), of the
residue ($r$), and of its constituent core ($c$) the residue, core and the
proton interactions with the target are computed in the $t\rho\rho$ and
$t\rho$ double- and single-folding approximations. A Gaussian, finite-range
nucleon-nucleon (NN) effective interaction is assumed,
\begin{eqnarray}
t_{jk}(r)=-\frac{\hbar v}{2}\sigma_{jk} \left(i+\alpha_{jk}\right) g
(\beta_{jk},r)\ ,
\end{eqnarray}
with $j,k=n,p$ and where $\sigma_{jk}$ and $\alpha_{jk}$ are the NN
total cross sections, taken from the parameterization of Ref.\ \cite{CG},
and the ratios of the real to the imaginary parts of their forward-scattering
amplitudes, taken from Ref.\ \cite{Ray}. Here $v$ is the particle-target
relative velocity and the Gaussian range parameters $\beta_{jk}$ are taken
to be 0.5 fm.

The weakly-bound $^{8}$B and $^9$C residue two-body wave functions were
calculated as $1p_{3/2}$ eigenstates with separation energies 0.137 MeV
and 1.3 MeV in a Woods-Saxon plus Coulomb potential of standard geometry
$r_0=1.25$ fm, $a_0=0.7$ fm and with a spin-orbit term of strength $V_{so}
=6.0$ MeV. The $\langle p,^{8}$B$|^{9}$C$\rangle$ and $\langle n,^{9}$C$|
^{10}$C$\rangle$ radial overlaps were likewise described by normalized
$1p_{3/2}$ Woods-Saxon eigenstates using the same geometry parameters
with separation energies $S_p$=1.3 MeV and $S_n$=21.3 MeV, respectively.
That is, we assume spectroscopic factors of 1.0 and calculate the cross
sections for one unit of single-particle strength.

For $^9$C$(-p)$, the calculated reaction cross section $\sigma_R$ between
the $^8$B residue and the $^9$Be target from the few-body and density-based
S-matrices are 851.8 mb and 856.1 mb. The smaller few-body model reaction
cross section is consistent with the higher-energy calculations in Refs.
\cite{ATT,AT} and the interpretations given in Refs. \cite{JG,NS2k}. The
corresponding $^9$C$(-p)$ single-particle cross sections using the few-body
(including break-up) and density-based (neglecting break-up) S-matrices are
61.25 mb and 61.47 mb, respectively.

For $^{10}$C$(-n)$, the corresponding few-body and density-based $^9$C
residue-target $\sigma_R$ are 866.2 mb and 868.0 mb. Again, the few-body
S-matrix reduces the reaction cross section, the difference being somewhat
smaller due to the increased proton separation energy from $^9$C of 1.3 MeV.
The $^{10}$C$(-n)$ single-particle cross sections when using
the few-body and density-based residue-target S-matrices are 21.44 mb and
21.51 mb, respectively.

Thus, as found with earlier higher-energy analyses \cite{AT,ATT}, these
consistent few-body model calculations indicate that the additional transparency
of the residue-target collisions, due to their extended wave functions (Fig.
1) - that would tend to increase the residue survival probability and
the removal cross section - is essentially balanced by the additional loss
of flux from the elastic channel due to break-up, that drives a reduced core
survival probability. These two competing aspects of the residue absorption
in the collision are not identified separately within the model calculations
discussed here. The overall effect upon the calculated nucleon-removal cross
section, being 0.3--0.4\%, is insignificant when compared to the precision
of typical, available experimental measurements and with other parameter
uncertainties and sensitivities in the model calculations presented.

\section{Summary comments}
We have compared model calculations of single-nucleon removal cross sections
for the $^9$C$(-p)$ and $^{10}$C$(-n)$ reactions when: (i) neglecting, and
(ii) including the break-up degrees of freedom of the weakly-bound $^8$B and
$^9$C reaction residues. Importantly, the calculations of methods (i) and (ii)
use the same parameters to describe these residues and their one-body density
distributions are computed consistently from the inputs used in their few-body
model description. The inclusion of break-up, through the eikonal elastic
S-matrix of the two-body residues with the target, reduces the calculated
reaction cross sections for the residue-target systems -- the consequence of
the granularity of the spatially-extended residues -- as has been documented
previously. We find that the inclusion of the break-up degree of freedom has
a very small effect on the calculated single-particle nucleon-removal cross
sections. For the representative $^9$C$(-p)$ and $^{10}$C$(-n)$ reactions
considered, chosen to involve the removal of both strongly- and weakly-bound
valence nucleons, the cross sections are found to be reduced by only 0.3--0.4\%.
We conclude that an explicit treatment of the break-up of weakly-bound residues
is unnecessary provided a realistic description of the residue density is used
in the no break-up, optical-limit calculations.

\bibliographystyle{apsrev}

\end{document}